\documentclass{article}
\usepackage{graphicx,epsfig,array}
\usepackage{amssymb,amsmath}
\usepackage{cite}

\begin{document}
   
\begin{center}
\large \bf
Solitary  waves in  one-dimensional pre-stressed lattice and its continual analog
\end{center}
\vspace*{3 mm}

\centerline{\bf Vsevolod Vladimirov$^\dagger$\footnote{Corresponding author: \texttt{vsevolod.vladimirov@gmail.com}}, Sergii Skurativskyi$^\ddagger$\footnote{ Corresponding author: \texttt{skurserg@gmail.com}}}

\vspace*{3 mm}

\begin{center}
\textit{$^\dagger$Faculty of Applied Mathematics, AGH University of Science
and Technology, \\Al. Mickiewicza 30, 30-059 Krakow, Poland}\\
\textit{$^\ddagger$Subbotin Institute of Geophysics, NAS of Ukraine, \\
Bohdan Khmelnytskyi str. 63-G, Kyiv, Ukraine}
\end{center}

\vspace*{3 mm}

\textit{Abstract:} One of the most interesting phenomena occuring in nonlinear media models
is the existence of wave patterns, such as kinks, solitons, compactons, peakons and
many others. There are known numerous nonlinear evolutionary PDEs, supporting
soliton (multi-soliton) and compacton traveling wave (TW) solutions. Unfortunately,
the vast majority of the models, with the exception of completely integrable ones, do
not enable to analyze the properties of solitary waves interaction using only
qualitative methods. Therefore it is instructive, when dealing with the non-integrable
PDEs, to combine the qualitative treatment with numerical simulations. In this report
we are going to present the results of studying compacton solutions in the continual
models for granular pre-stressed chains. The model is shown to possess a pair of
compacton TW solutions which are the bright and dark compactons. First we consider
the stability properties of the compacton solutions and show that both the bright and
the dark compactons pass the stability test. Next we analyze the dynamics of the
compactons, simulating numerically the temporal evolution of a single compacton, a
well as the interection of pairs of compactons, including bright-bright, dark-dark and
bright-dark pairs. To be able to simulate the evolutin of interacting compactons, we
have modified the numerical scheme built by J. de Frutos, M. A. Lopez-Marcos, and J.
M. Sanz-Serna. Results of simulations are compared with that of evolution of
corresponding impulse in the granular pre-stressed chain.



\section*{Introduction} This paper deals with some nonlinear evolutionary PDEs associated with dynamics of one-dimensional chains of pre-stressed granules. Since  Nesterenko's pioneering works, \cite{Nester_83,Nester_94,Nester_02}, propagation of pulses in such  media has been a subject of a great number of experimental studies and numerical works.  We consider a nonlinear evolutionary PDE  associated with an ODE systems describing the interaction of the adjacent elements of the chain with  the forces depending on the relative displacement of the centers of mass of the granules.   The PDE in question is obtained by means of the passage to the continuum limit, followed by the formal multi-scale decomposition.

We  perform in this paper qualitative and numerical  study of dark and bright compacton traveling wave solutions, supported by the PDE, paying attention to the   stability and dynamical features of the compactons' solutions.  The paper is arranged as follows. In section 2 we introduce  the continual analog of the granular pre-stressed media with the specific interaction allowing for the propagation of  the wave of compression as well as the wave of rarefaction.  In section 3 we construct the Hamiltonian representations for  the model, and show that the compacton traveling wave solutions, satisfying factorised equations, fulfill necessary conditions of extrema for some Lagrange functionals.  Next we perform  stability tests  for compacton solutions, basing on the approach developed in \cite{Derrick_64, Zakharov_86,Karpman_95}, and show that both the dark and the bright compactons pass the stability test. The results of qualitative analysis are backed and partly supplemented by the numerical study. Numerical simulations show that the compacton solutions  completely reestablish their shapes after the mutual collisions. 
Finally, in section 4 we present the results of numerical simulation of the Cauchy problem for discrete chains and compare the results of the numerical simulation  with the analogous simulations performed within the continual analog.

\section{Evolutionary PDEs associated with the granular prestressed chains}

Unusual features of the solitons associated with the celebrated Korteveg-de Vries (KdV) equation, as well as other completely integrable  models \cite{Dodd}, are often prescribed to the existence of higher symmetries and (or) infinite set of conservation laws. However, there are known non-integrable equations possessing the localised TW solutions with quite similar  features. As a well-known example, the so called $K(m,\,n)$ hierarchy \cite{Hyman-Rosenau} can be presented:
\begin{equation}\label{Kmn}
K(m,\,n)=u_t+\left( u^m \right)_x+ \left(u^n\right)_{xxx}=0, \qquad m\geq 2,\,\,\,n\geq 2.
\end{equation}
Members of this hierarchy are not completely integrable at least for the generic values of the parameters $m$, $n$ \cite{Vodova} and yet possess the compactly-supported TW solutions demonstrating the solitonic features \cite{Hyman-Rosenau, frutos}.

The $K(m,\,n)$ family was introduced in years 90th of the XX century as a formal generalisation of the KdV hierarchy, without referring to its physical context. Earlier V.F. Nesterenko \cite{Nester_83} considered the dynamics of a chain of preloaded granules described by the following ODE system:
\begin{equation}\label{DS_Nest}
\ddot Q_k(t)=F(Q_{k-1}-Q_{k})-F(Q_{k}-Q_{k+1}),   \qquad       k\,\in \left \{0, \pm\,1, \pm\,2,....   \right\}
\end{equation}
where
$Q_k(t)$ is the displacement of granule $k$ centre of mass from its equilibrium position,
\begin{equation}\label{force}
F(z)=A\,z^n, \qquad n>1.
\end{equation}
A passage to the continual analog of the above discrete model is attained  by the substitution:
\begin{equation}\label{Contanalog}
Q_k(t)=u(t,\,k\,a) \equiv u(t,\,x),
\end{equation}
where $a$ is the average distance between granules.
Inserting this formula, together with the identities
\begin{equation}\label{ContShift}
Q_{k\pm\,1}=u(t,\,x\pm\,a)=e^{\pm a\,D_x}u(t,\,x)=\sum_{j=0}^{n+3} \frac{(\pm\,a)^j}{j!}\frac{\partial^j}{\partial\,x^j}u(t,\,x)+O\left(|a|^{n+4}\right),
\end{equation}
into (\ref{DS_Nest}) and dropping out terms of the order $O(|a|^{n+4})$ and higher, we get the equation:
\begin{equation}\label{PDE2_1}
u_{tt}=-C\,\left\{\left(-u_x  \right)^n+\beta\,\left(-u_x  \right)^\frac{n-1}{2} \left[\left(-u_x  \right)^\frac{n+1}{2}  \right]_{xx}   \right\}_x,
\end{equation}
where
\[
C=A\,a^{n+1}, \qquad \beta=\frac{n\,a^2}{6(n+1)}.
\]
Differentiating the above equation with respect to $x$ and employing the new variable $S=\left(-u_x  \right)$, we obtain the Nesterenko's equation \cite{Nester_02}:
\begin{equation}\label{EqNest}
S_{tt}=C\,\left\{S^n+\beta\,S^\frac{n-1}{2} \left[S^\frac{n+1}{2}  \right]_{xx}   \right\}_{xx}.
\end{equation}
Eq. (\ref{EqNest}) describes dynamics of strongly preloaded media in which the propagation of acoustic waves is impossible (the effect of "sonic vacuum"). Nesterenko  had shown \cite{Nester_02} that this equation possesses a one paremeter family of compacton TW solutions describing the propagation of the waves of compression. Unfortunately, he did not pay much attention to the investigation of their dynamical features. Our preliminary study show that the compacton solutions supported by Eq. (\ref{EqNest}) are unstable. This situation is absolutely analogous to that with processing the celebrated Fermi-Pasta-Ulam problem \cite{Dodd} which takes the form of  system 
(\ref{DS_Nest}) in which the interaction force has the  form $F(z)=A \,z^2+B\, z$ with $|A|=O(|B|)$. Passing to the continual analog by means of the substitution (\ref{Contanalog}) one obtains the Boussinesq equation, possessing unstable soliton-like solutions. The famous KdV equation is extracted from the Boussinesq equation by means the multi-scale decomposition \cite{Dodd}.

Our proceedings to the "proper" compacton-supporting equation is following. We start from the  discrete system (\ref{DS_Nest}) in which the interaction force has the form
\begin{equation}\label{Forceint}
F(z)=A\,\,z^n+B\,z.
\end{equation}
In addition, we assume that  $B=\gamma\,a^{n+3}$, $|\gamma|=O(|A|)$. Making in the formula (\ref{DS_Nest}) the substitution (\ref{Contanalog}), (\ref{ContShift}) we get the equation
\begin{equation}\label{PDE2}
u_{tt}=-C\,\left\{\left(-u_x  \right)^n+\beta\,\left(-u_x  \right)^\frac{n-1}{2} \left[\left(-u_x  \right)^\frac{n+1}{2}  \right]_{xx}   \right\}_x+\gamma\,a^{n+3}\,\left(-u_x  \right)_{x}.
\end{equation}
Differentiating the above equation with respect to $x$ and introducing the new variable $S=\left(-u_x  \right)$,  we obtain the following equation:
\begin{equation}\label{EqNest2}
S_{tt}=C\,\left\{S^n+\beta\,S^\frac{n-1}{2} \left[S^\frac{n+1}{2}  \right]_{xx}   \right\}_{xx}+\gamma\,a^{n+3}\,S_{xx}.
\end{equation}

Now we use a series of scaling transformations. Employment of  the scaling
$
\tau =\sqrt{\gamma\,a^{n+3}}\,t,
$
enables to  rewrite the above equation in the form:
\[
S_{\tau\,\tau}=\frac{C}{\gamma\,a^{n+3}}\,\left\{S^n+\beta\,S^\frac{n-1}{2} \left[S^\frac{n+1}{2}  \right]_{xx}   \right\}_{xx}+S_{xx}.
\]
Next the transformation
$
\bar T=\frac{1}{2} a^q \tau, \quad \xi=a^p (x-\tau), \quad S=a^r\,W
$
is used. If, for example, we make a choice
$
q=1, \quad p=-1, \, r=5,
$
then the higher order coefficient $O(a^2)$ will stand at the term with the second derivative with respect to $\bar T$. So, dropping out the terms proportional to $O(a^2)$, we obtain, after the integration with respect to $\xi$, the equation:
\[
W_{\bar T}+\frac{A}{\gamma}\left\{W^n+\frac{n}{6\,(n+1)} W^\frac{n-1}{2} \left[W^\frac{n+1}{2}  \right]_{\xi\,\xi}  \right\}_{\xi}=0.
\]
The scaling
$
 T=\frac{A}{\gamma}\,L \, \bar T, \quad X=L\,\xi, \quad L=\sqrt{\frac{6 (n+1)}{n}}
$
leads us finally to the target equation:
\begin{equation}\label{PDE3}
W_{ T}+\left\{W^n+ W^\frac{n-1}{2} \left[W^\frac{n+1}{2}  \right]_{XX}  \right\}_{X}=0.
\end{equation}

Description of  waves of rerefaction in the case $n=2\,k$ requires the following modification of the interaction force:
\begin{equation}\label{Forcerar}
F(z)=-A z^{2\,k}+B\,z
\end{equation}
(for $n=2\,k+1$ the formula (\ref{Forceint}) describes automatically both wave of compression and raferaction). Applying the above machinery to the system (\ref{DS_Nest}) with the interaction (\ref{Forcerar}), one can get the equation
\begin{equation}\label{PDErar}
W_T-\left\{W^n+ W^\frac{n-1}{2} \left[W^\frac{n+1}{2}  \right]_{XX}  \right\}_{X}=0, \quad n=2\,k.
\end{equation}
Thus, for $n=2\,k$ the universal equation describing waves of compression and rarefaction can be presented in the form
\begin{equation}\label{PDEComprar}
W_{ T}+\mathrm{sgn}(W)\,\left\{W^n+ W^\frac{n-1}{2} \left[W^\frac{n+1}{2}  \right]_{XX}  \right\}_{X}=0.
\end{equation}

Let us note in conclusion that the equations (\ref{PDE3}) - (\ref{PDEComprar}) are obtained by formal employment of the multiscale decomposition method, which cannot be justified in our case because of negativeness of the index $p$. Nevertheless further investigations of these equations is still of interest because they occur to possess a  set of compacton  solutions demonstrating interesting dynamical features. As will be shown below, these solutions qualitatively correctly describe propagation of short impulses in the chain of prestressed blocks.

\section{Compacton solutions and stability tests}


Let us consider the pair of equations (\ref{PDE3}), (\ref{PDErar}), which can be expressed in the common form
\begin{equation}\label{Comprar}
W_{ T}+\epsilon\,\left\{W^n+ W^\frac{n-1}{2} \left[W^\frac{n+1}{2}  \right]_{XX}  \right\}_{X}=0, \quad \epsilon=\pm\,1.
\end{equation}   
Since we are interested in the TW solutions $W=W(z)\equiv W(X-c\,T)$,  it is instructive
to make a passage to the traveling wave coordinates  $T\to T, \quad X\to z=X-c\,T$. Performing this change,  we get:
\begin{equation}\label{PDE3C}
W_{ T}-c\,W_{ z}+\epsilon\,\left\{W^n+ W^\frac{n-1}{2} \left[W^\frac{n+1}{2}  \right]_{zz}  \right\}_{z}=0.
\end{equation}

 Below we formulate several statements, which are easily verified by direct inspection.

{\bf Statement 1.} {\it
Eq. (\ref{PDE3C})
admits the following  representation
\begin{equation}\label{Hamrepr2}
\frac{\partial}{\partial T} W = \frac{\partial}{\partial\,z}\,
{\delta\,\left( \epsilon\,H+c \,Q\right)}/{\delta\,W},
\end{equation}
where
\[
H=\int\left[\frac{n+1}{4}W^{n-1} \,W_X^{2}-\frac{1}{n+1}\,W^{n+1}\right]\,d\,z,
\qquad
Q=\frac{1}{2}\int W^2\,d\,z.
\]
}

\vspace{3mm}

{\bf Statement 2.} {\it
The functionals $H$, $Q$ are conserved in time.
}

\vspace{3mm}

{\bf Statement 3.} {\it
Consider the following functions:
\begin{equation}\label{comp2a}
W_c^\epsilon(z)=\epsilon\,W_c(z)=\left\{\begin{array}{l}\epsilon\, M \cos^{\gamma}{\left(B\,z\right)},\quad \mathrm{if}\,\,|K\,z|<\frac{\pi}{2}, \\
0 \qquad\qquad \mathrm{elsewhere},\\
\end{array}\right.
\end{equation}
where
\[
 M=\left[\frac{c (n+1)}{2}  \right]^{\frac{1}{n-1}}, \qquad K=\frac{n-1}{n+1},  \qquad \gamma=\frac{2}{n-1}.
\]
If $n=2\,k+1,\quad k\,\in\,\mathrm{N},$ then the functions $W_c^\pm(z)$ are the generalized solutions to the equation
\begin{equation}\label{variat2D}
\delta\,\left( H+c \,Q\right)/\delta\,W|_{W=W_c^\pm} =0.
\end{equation}

 If $n=2\,k,\quad k\,\in\,\mathrm{N},$ then the function $W_c^\epsilon(z)$ satisfies the equation
\begin{equation}\label{variateps}
\delta\,\left( \epsilon\,H+c \,Q\right)/\delta\,W|_{W=W_c^\epsilon} =0. 
\end{equation}
 }

So, the TW solutions  (\ref{comp2a}) are  the critical points of either the Lagrange functional $\Lambda[\beta]=\left(H+\beta\,Q\right)$ (case $n=2\,k+2$) or $\Lambda^\epsilon[\beta]=\left(\epsilon\,H+\beta\,Q\right)$ (case $n=2\,k$) with the common Lagrange multiplier $\beta=c$. A necessary and sufficient condition for $\Lambda[\beta]$ ($\Lambda^\epsilon[\beta]$) to attain the minimum on the compacton solution can be  formulated in terms of the positiveness of the second variation of the corresponding functional, which, in turn, guarantees the orbital stability of the TW solution, \cite{KaPromis}. Here we do not touch upon the problem of strict estimating the signs of the second variations. Instead of this, we follow the approach suggested in \cite{Derrick_64,Zakharov_86,Karpman_95}, which enables to test  a mere possibility of the local minimum appearance  on a selected sets of perturbations of TW solutions.

Let us consider a family of perturbations
\begin{equation}\label{scaling1A}
W_c^\epsilon(z) \rightarrow \lambda^\alpha W_c^\epsilon(\lambda\,z).
\end{equation}
By choosing $\alpha=1/2$ we guarantee that
\begin{equation}\label{Qinv}
Q[\lambda]=\frac{1}{2}\int_{-\pi/2}^{\pi/2}{\left[\lambda^\frac{1}{2}\,W_c^\epsilon(\lambda\,z)\right]^2\,d\,z}=Q[1].
\end{equation}
Imposing this condition, we reject the ''longitudinal'' perturbations, associated with symmetry $T_\gamma\left[W_c^\epsilon(z)\right]=W_c^\epsilon(z+\gamma)$. Indeed, since the equations (\ref{variat2D}), (\ref{variateps}) are invariant under the shift $z\,\to\, z+\gamma$, then  $T_\gamma W_c^\epsilon(z)$ belongs to the set of solutions as well, while formally the transformation $W_c^\epsilon(z) \rightarrow W_c^\epsilon(z+\gamma)$ can be treated as a perturbation. In order to exclude the perturbations of this sort, the orthogonality condition  is posed. Introducing the representation for the perturbed solution
\[
W_c^\epsilon(z)[\lambda]=W_c^\epsilon(z)[1]+v(z),
\]
and using the condition (\ref{Qinv}), we get
\[
0=Q[\lambda]-Q[1]=\int_{-\pi/2}^{\pi/2} W_c^\epsilon(z)\,v(z)\,d\,z+O\left(||v(z)||^2\right).
\]

For $\alpha=1/2$, we get the following functions to be tested:
\begin{equation}\label{flam1Q}
\Lambda^\nu[\lambda]=(\nu\,H+c \,Q)[\lambda]=\nu\,\left\{\lambda^{\frac{n+3}{2}}\, I_{n}^\epsilon-\lambda^{\frac{n-1}{2}}\, J_{n}^\epsilon\right\}+c\,Q,
\end{equation}
where
\[
 I_{n}^\epsilon=\frac{n+1}{4}\int_{-\pi/2}^{\pi/2}\,\left[W_c^\epsilon\right]^{n-1}(\tau)\,\left[W_c^\epsilon\right]_\tau^2(\tau)\,d\,\tau, \quad
 J_{n}^\epsilon=\frac{1}{n+1}\int_{-\pi/2}^{\pi/2}\,\left[W_c^\epsilon\right]^{n+1}(\tau)\,d\,\tau,
\]
\[
\nu=\epsilon^{n+1}=\left\{\begin{array}{l} +1 \qquad \mathrm{when}\,\,n=2\,k+1, \\
\quad \epsilon\qquad \mathrm{when}\,\,n=2\,k. \\
\end{array}\right.
\]

If the functional $\Lambda^\nu=\nu\,H+c \,Q$ attains the extremal value on the compacton solution, then the function $\Lambda^\nu[\lambda]$ has the corresponding extremum in the point $\lambda=1$. The verification of this property is used  as a test.

A necessary condition of the extremum
$
\frac{d}{d\,\lambda}\Lambda^\nu[\lambda]\Bigl|_{\lambda=1}=0
$
gives us the equality:
\begin{equation}\label{In2}
 I_n^\epsilon=\frac{n-1}{n+3} J_n^\epsilon.
\end{equation}
Using (\ref{In2}), we can easily get convinced that 
\[
\frac{d^2}{d\,\lambda^2}\Lambda^\nu[\lambda]\Bigl|_{\lambda=1}=\nu\,\frac{n-1}{2} J_n^\epsilon\,=
\epsilon^{2(k+1)}\frac{n-1}{2(n+1)}\int\,\left[W_c\right]^{n+1}(\tau)\,d\,\tau\,>\,0
\]
for both $n=2\,k+1$ and $n=2\,k.$

So the generalized solutions (\ref{comp2a})  pass the test for stability.

Further information about the properties of the compacton solutions deliver the numerical simulations.

\section{Numerical simulations of compactons' dynamics}

\begin{figure}[h]
\begin{center}
\includegraphics[width=5.5 cm, height=5
cm]{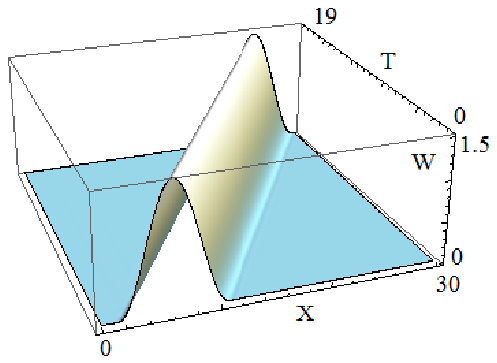}\hspace{0.5 cm}
\includegraphics[width=0.5\linewidth]{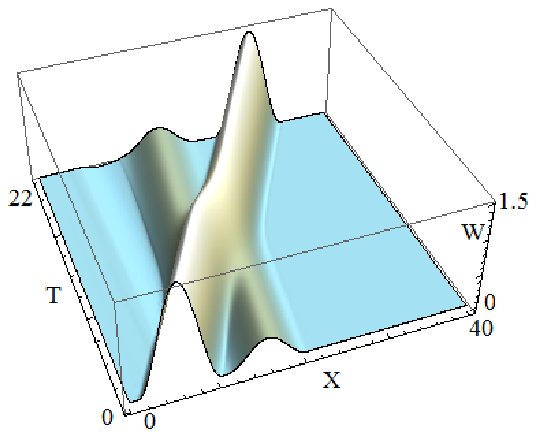}\\
a) \hspace{6 cm} b)\\
\end{center}
\caption{The movement of single compacton with the velocity $D=1$
to the right (left panel). The movement of two bright compactons with
the velocities  $D=1$ and $D=1/4$, respectively (right panel).
}\label{fig:1}
\end{figure}

\begin{figure}
\begin{center}
\includegraphics[width=5.5 cm, height=5
cm]{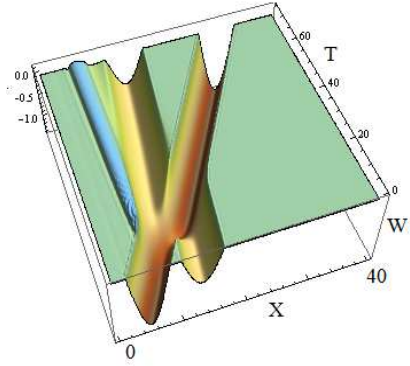}
\end{center}
\caption{The movement of two dark-compactons  with the
velocities $D=1$ and $D=1/4$, respectively.}\label{fig:2}
\end{figure}
The solitary waves' dynamics is studied by means of 
direct numerical simulation, based on the finite-difference scheme. 
To derive the finite-difference scheme, e.g., for the model equation (\ref{PDE3}), we modify the scheme presented in \cite{frutos}.  
In accordance with the metodology proposed in this paper, we introduce the artificial viscosity by adding the term $\varepsilon
W_{4x}$, where $\varepsilon$ is a small parameter. Thus, instead of
 (\ref{PDE3}) we have in the case $n=3$ the following equation:

\begin{equation}
W_t+\left\{W^3\right\}_x+\left\{W\left[W^2\right]_{xx}\right\}_x+\varepsilon
W_{4x}=0.
\end{equation}
Let us approximate the spatial derivatives as follows
\begin{equation}\label{descrete}
\begin{split}
\frac{1}{120}(\dot W_{j-2}+26\dot W_{j-1}+66\dot W_j+26\dot
W_{j+1}+\dot W_{j+2})+
\\+\frac{1}{24h}(-W_{j-2}^3-10 W_{j-1}^3+10 W_{j+1}^3+W_{j+2}^3)+\\
+\frac{1}{24h}(-L_{j-2}-10 L_{j-1}+10 L_{j+1}+L_{j+2})+ \\
+\varepsilon \frac{1}{h^4}(W_{j-2}-4 W_{j-1}+6W_j
-4W_{j+1}+W_{j+2})=0,
\end{split}
\end{equation}
where $L_j=W_j\frac{W_{j-2}^2-2W_j^2+W_{j+2}^2}{h^2}$ .

To integrate the system (\ref{descrete}) in time, we use the midpoint
method. According to this method, the quantities $W_j$ and $\dot
W_j$ are presented in the form
$$ W_j\rightarrow \frac{W_j^{n+1}+W_j^{n}}{2}, \dot W_j\rightarrow
\frac{W_j^{n+1}-W_j^{n}}{\tau}.$$ The resulting nonlinear
algebraic system with respect to  $W_j^{n+1}$ can be solved by
iterative methods.

We test the scheme (\ref{descrete}) by considering the movement of a
singe compacton.  Assume that  the model's parameters $D_1=1$,
$s_0=5.5$ and scheme's parameters $N=600$, $h=30/N$, $\tau=0.01$,
$\varepsilon=10^{-3}$ are fixed. The application of the scheme
(\ref{descrete}) gives us the figure \ref{fig:1}a.

To study the interaction of two bright compactons, we combine  the
previous compacton and another one with the lower amplitude taking
the velocity $D_2=D_1/4$ and $s_0=15.5$. The result of modelling
is presented in fig.~ \ref{fig:1}b. The interaction of two dark compactons has the similar properties and is depicted in fig.~ \ref{fig:2}.

\begin{figure}
\begin{center}
\includegraphics[totalheight=1.4 in]{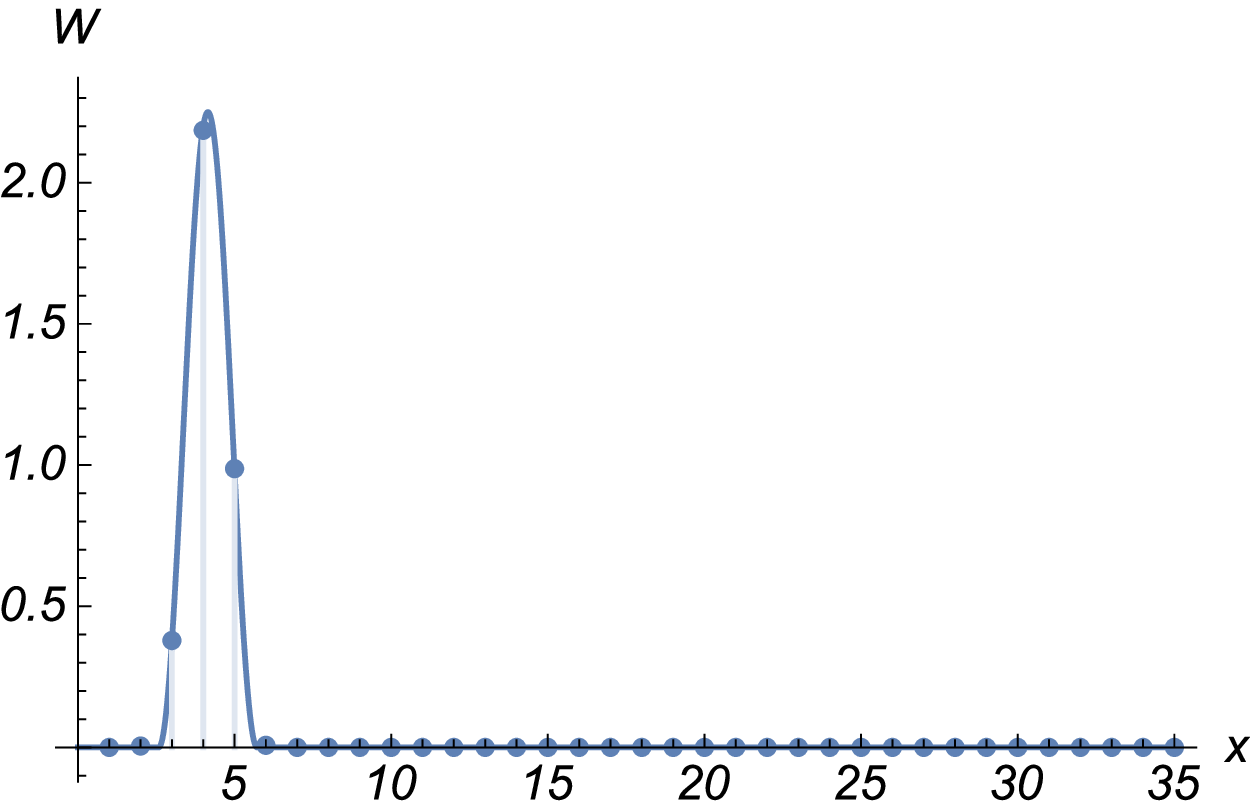}
\includegraphics[totalheight=1.4 in]{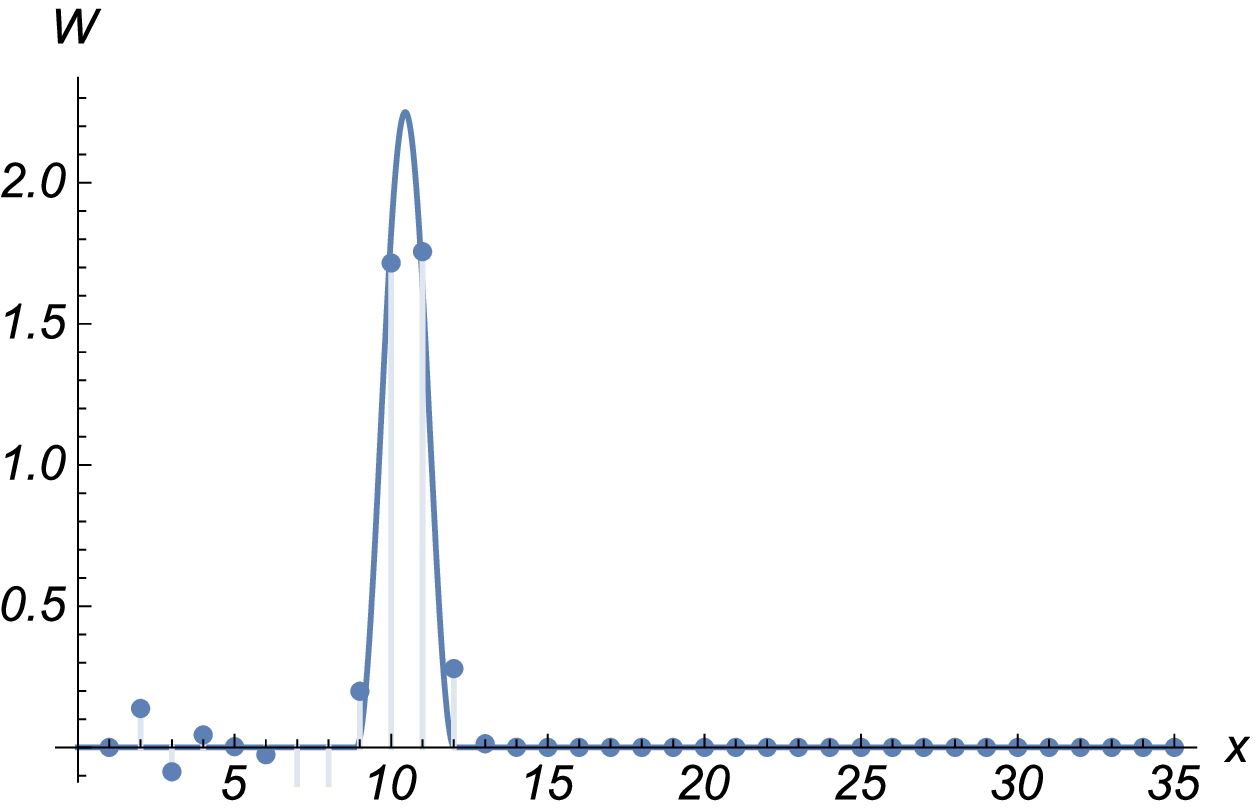}
\\
\includegraphics[totalheight=1.4 in]{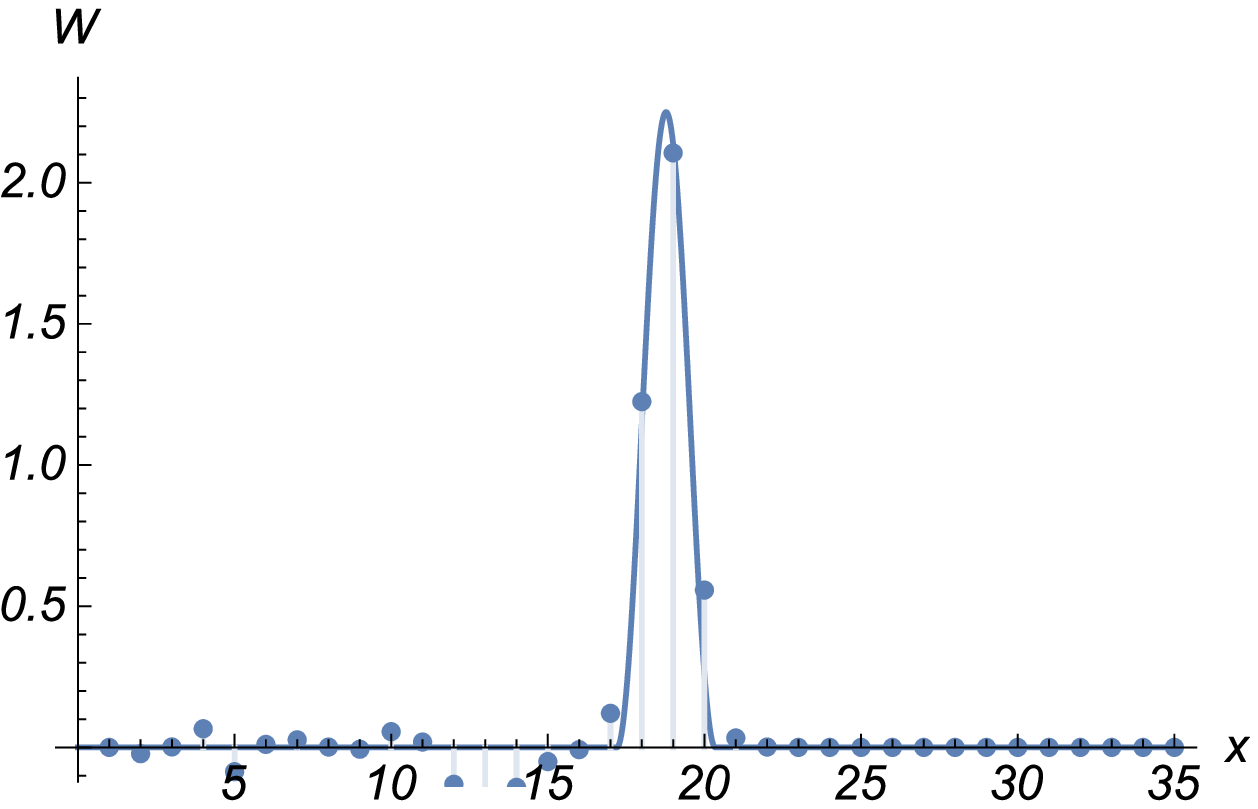}
\includegraphics[totalheight=1.4 in]{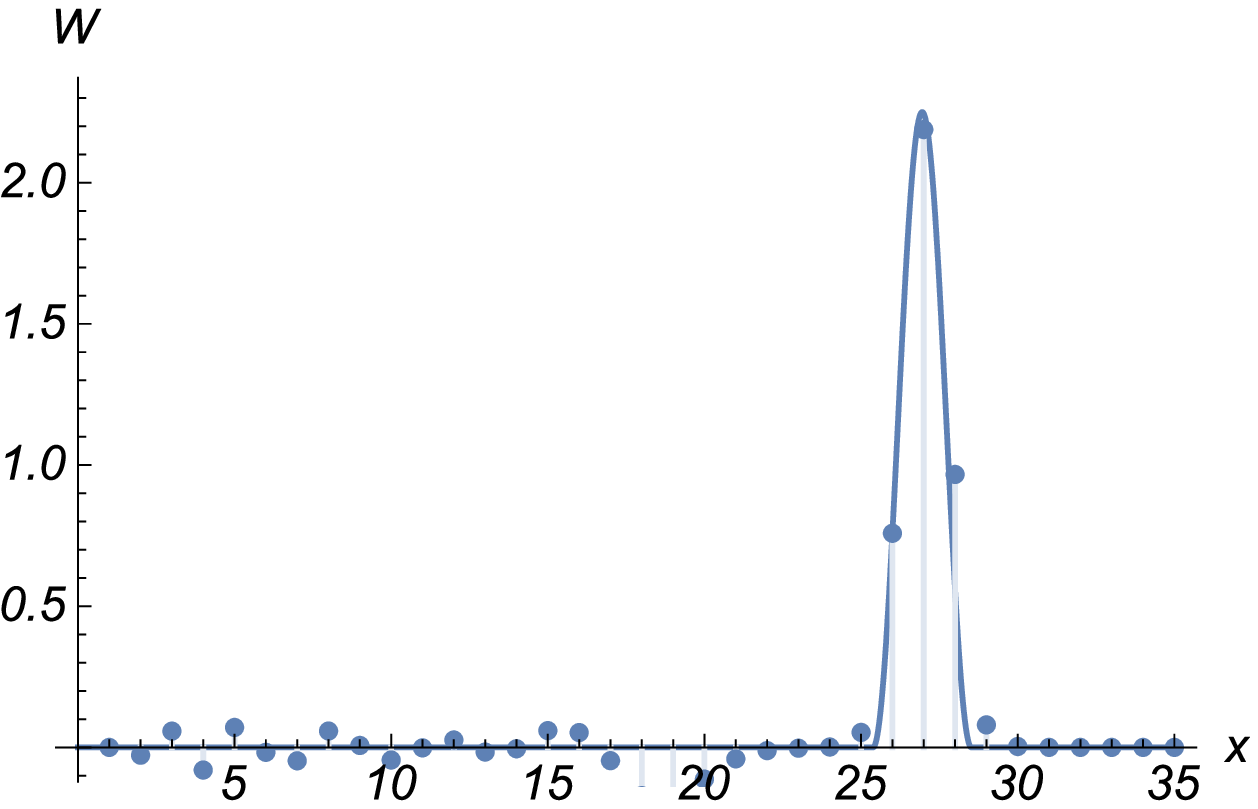}
\caption{Evolution of the initial perturbation in the granular media (marked with dots) on the background of the corresponding evolution of the  compacton (marked with solid lines) upper row: left: $t=0$ right: $t=4$; lower row: left: $t=9$; right: $t=14$
}\label{Fig:singcomp} \end{center}
\end{figure}

\section{Comparison of numerical evolution of compactons with the numerical solution of the granular media, subjected to similar initial conditions}

\begin{figure}
\begin{center}
\includegraphics[totalheight=1.4 in]{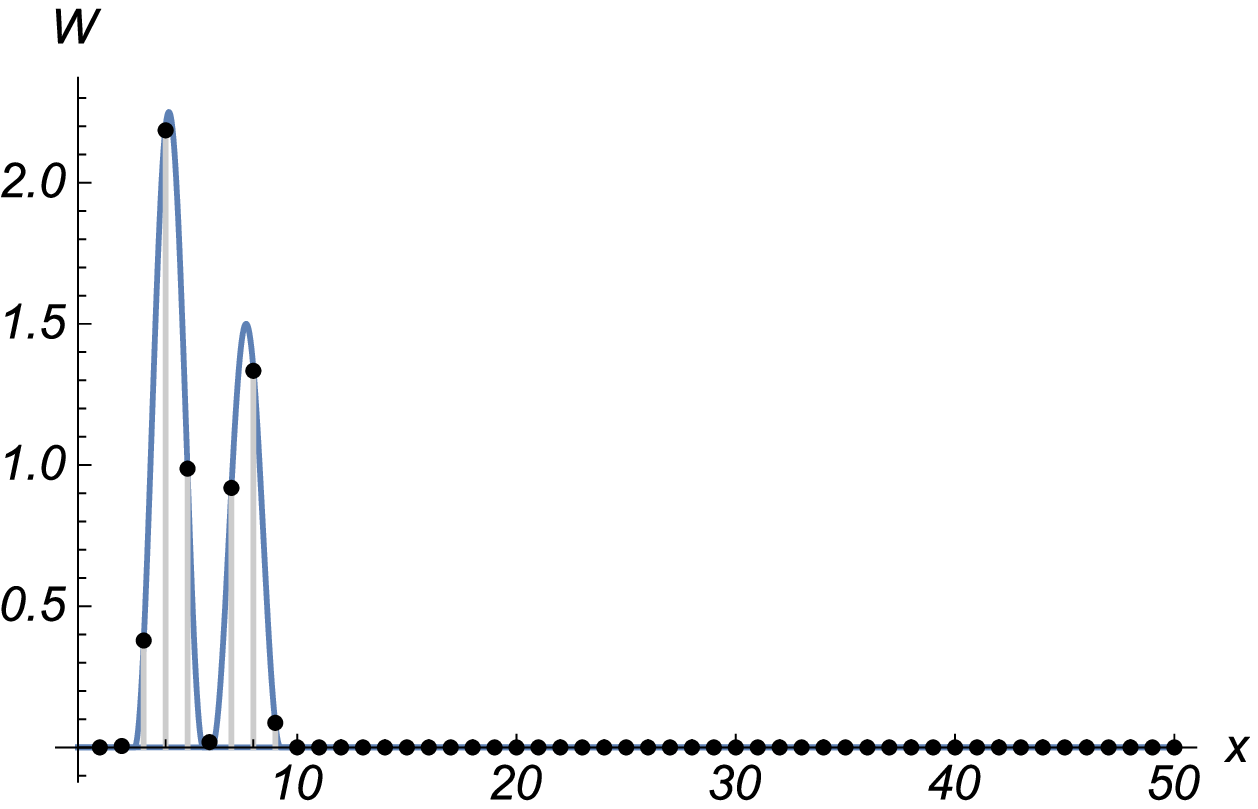}
\includegraphics[totalheight=1.4 in]{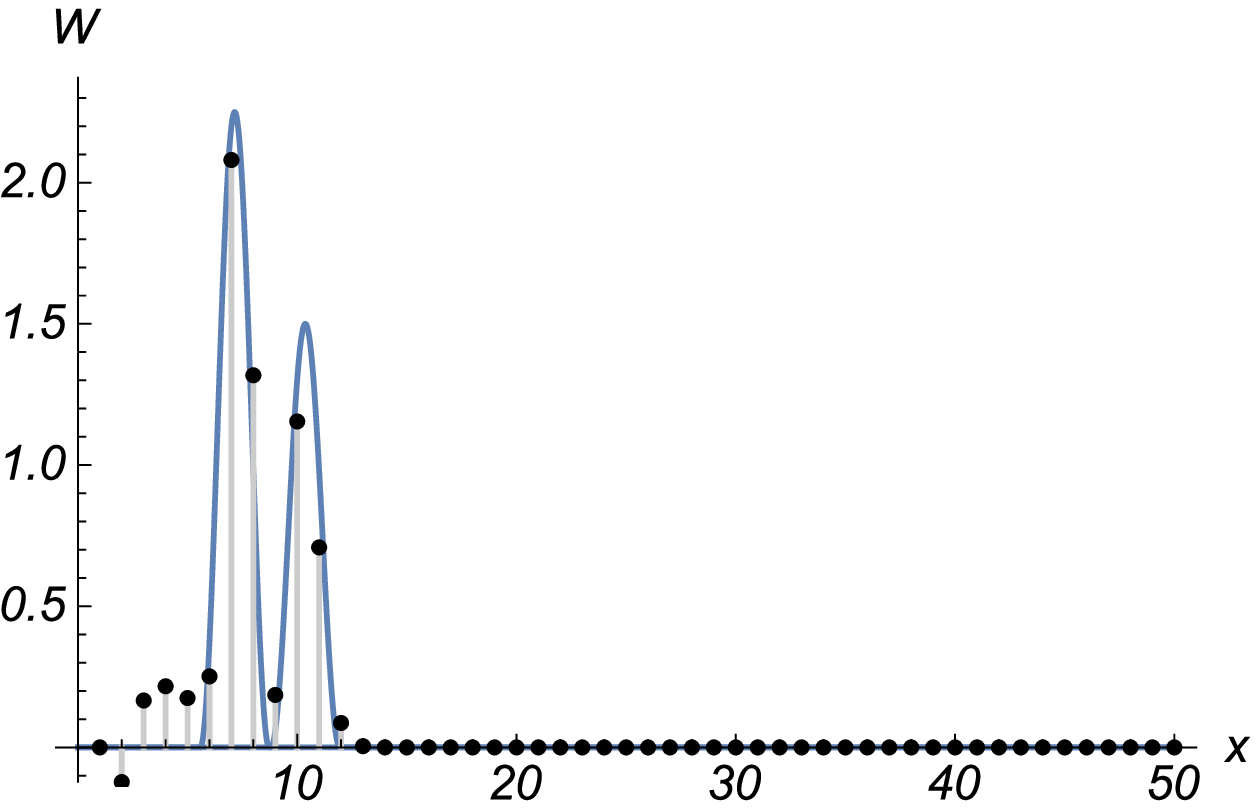}
\\
\includegraphics[totalheight=1.4 in]{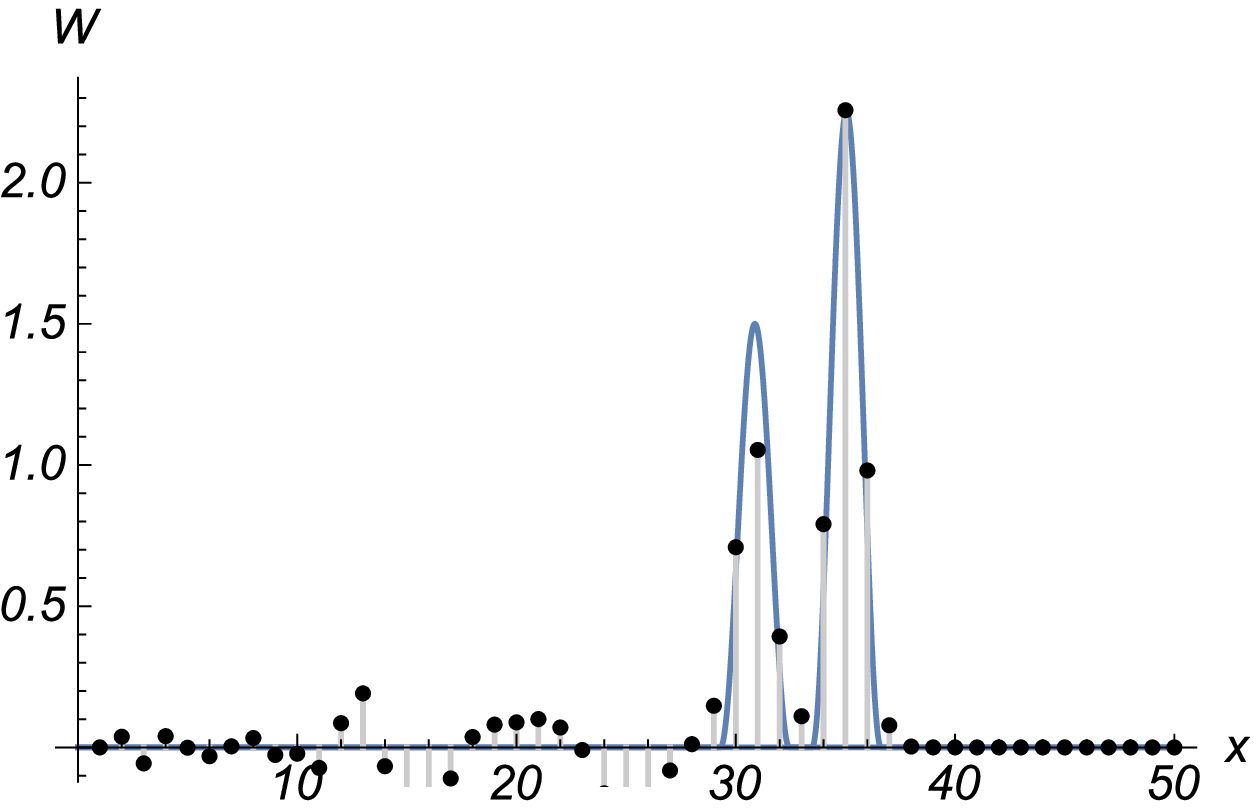}
\includegraphics[totalheight=1.4 in]{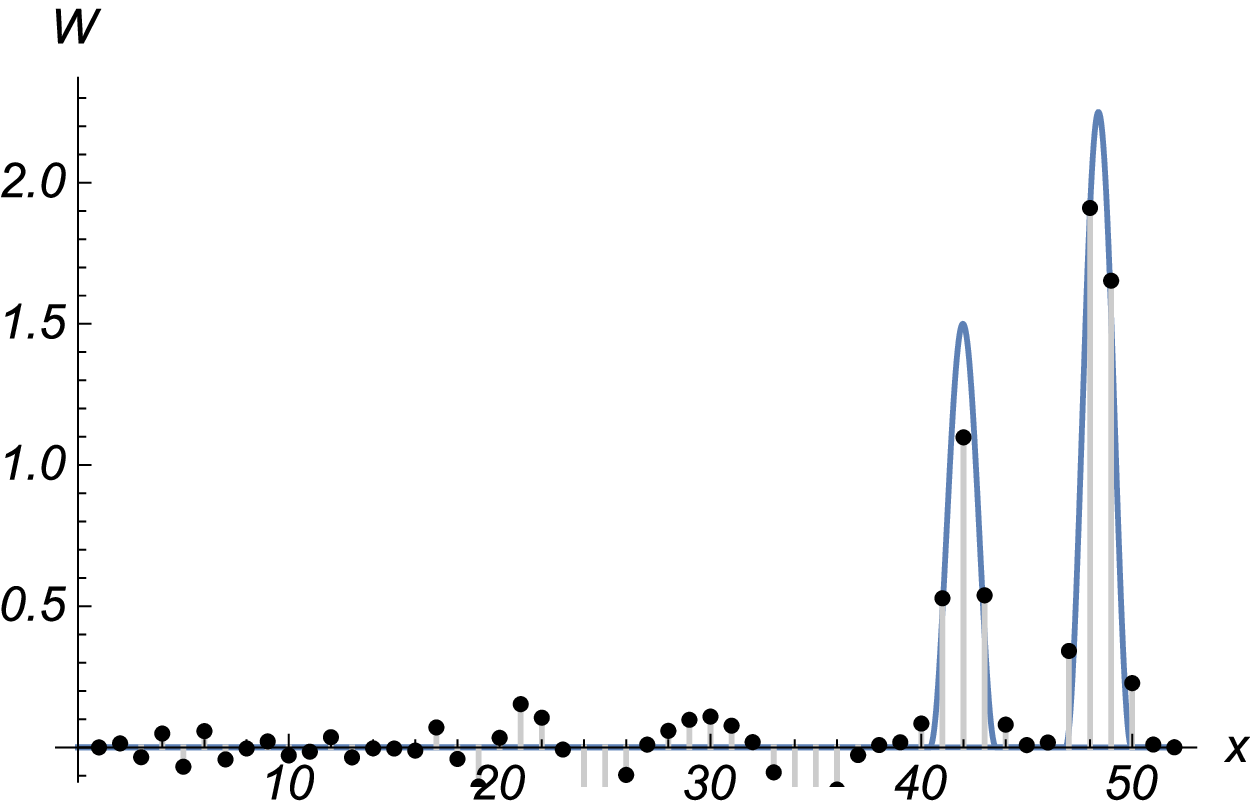}
\caption{Evolution of two initially separated compacton perturbation in the granular media (marked with dots) on the background of the numerical solution of the continual model with the same initial data  (marked with solid lines) upper row: left: $t=0$ right: $t=3$; lower row: left: $t=21.5$; right: $t=31$
}\label{Fig:collisions} \end{center}
\end{figure}

We've performed the comparison of the evolution of  compacton solutions  with {corresponding} solutions of the finite (but long enough) discrete system. The discrete analogs to the field ${S(t,\,x)=-\frac{\partial\,u(t,\,x)}{\partial\,x}}$
  are the  ''{stresses}'' $R_k=Q_{k-1}-Q_k$,  satisfying the system  
\[
\ddot R_1(t)=0, \quad
\ddot R_k(t)=A\left[ R_{k-1}^n-2\,R_k^n+R_{k+1}^n \right], \quad
\ddot R_m(t)=0
\]
with the initial conditions
\[
R_k(0) = \begin{cases}   M \cos^\gamma [ B a k-I]  & \mbox{if } |B a k-I|<\pi/2  \\ 
0 & \mbox{if }  otherwise,  \end{cases}
\]
and
\[
\dot R_k(0) = \begin{cases}   -M c \gamma \cos^{\gamma-1} [ B a k-I]\sin[ B a k- I]  &\mbox{if } |B \,a\,k-I|<\pi/2  \\ 
0 & \mbox{if }  otherwise,  \end{cases}
\] 
where $I$ is a constant phase,  $ k=2,3,....,m-1$. The result of comparison for a single compacton is shown in fig.~\ref{Fig:singcomp}. In fig.~\ref{Fig:collisions} evolution of two initially separated compactons is shown for both continual and discrete models.

\section{Conclusion and discussion}

In this paper compacton solutions are studied, supported by the continual analogs of  the dynamical systems describing one-dimensional chains of prestressed particles. The equation (\ref{EqNest2}) obtained without resorting to  the method of multi-scaled decomposition possesses the compacton solutions which fail to pass the stability test. Numerical experiments show that the compacton solutions are destroyed in a very short time.

Contrary, the equations  (\ref{PDE3}), (\ref{PDErar}) which are obtained with the help of formal multi scale decomposition,   possess families  of bright and dark  compacton solutions, correspondingly, which occur to be stable. This is backed both by the stability test and results of the numerical simulations.

A characteristic feature of the equation (\ref{PDE3}) connected with the decomposition used during its derivation is that it describes a processes with the "long" temporal and  "short" spatial scales. So it is rather questionable if this equation can adequately describe a localised pulse propagation in a discrete media in which the characteristic sizes of the particles are comparable with  compacton's width $\Delta x$. In fact, making the backward transformations $X\rightarrow\xi\rightarrow x$ we get the following formula for the width of the compacton solution (\ref{comp2a}) in the initial coordinate:
\[
\Delta x= \pi\,a\,\sqrt{\frac{n (n+1) }{  6 (n-1)^2}}.
\]
For $n=3/2$, corresponding to the Hertzian force between spherical particles, we get $\Delta x\approx 4.96\,a$. It is then curious to  know that the same results for the particles with the spherical geometry were obtained during the numerical work,  and experimental studies \cite{Nester_83, Nester_85,Nester_94,Nester_95,Vengr}.

Let us remark in conclusion that some of the present results, in particular those concerning the stability study, are only preliminary. The full investigations of stability of compacton solutions supported by (\ref{PDE3})-(\ref{PDEComprar}) will be published elsewhere.

\bibliographystyle{acm} 
\bibliography{DSTA15_VSAN_SKUR} 	

\begin{thebibliography}{10}

\bibitem{frutos}
{\sc de~Frutos, J., Lopez-Marcos, M.~A., and Sanz-Serna, J.~M.}
\newblock A finite-difference scheme for the k(2,2) compacton equation.
\newblock {\em J. Comput. Phys.}, 120 (1995), 248--252.

\bibitem{Derrick_64}
{\sc Derrick, G.~H.}
\newblock Comments on nonlinear wave equations as models for elementary
  particles.
\newblock {\em J. Math. Phys.}, 5 (1964), 1252.

\bibitem{Dodd}
{\sc Dodd, R., Eilbeck, J., Gibbon, J.~D., and Morris, H.}
\newblock {\em Solitons and Nonlinear Wave Equations}.
\newblock Academic Press, London, 1984.

\bibitem{Hyman-Rosenau}
{\sc Hyman, J., and Rosenau, P.}
\newblock Compactons: solitons with finite wavelelngth.
\newblock {\em Phys. Rev. Lett.}, 70 (1993), 564.

\bibitem{KaPromis}
{\sc Kapitula, T., and Promislov, K.}
\newblock {\em Spectral and Dynamical Stability of Nonlinear Waves}.
\newblock Springer-Verlag, New York, 2013.

\bibitem{Karpman_95}
{\sc Karpman, V.}
\newblock Stabilization of soliton instabilities by higher order dispersion:
  Kdv-type equations.
\newblock {\em Phys. Lett.}, A210 (1996), 77.

\bibitem{Zakharov_86}
{\sc Kuznetsov, E.~A., Rubenchik, A.~M., and Zakharov, V.}
\newblock Soliton stability in plasmas and hydrodynamics.
\newblock {\em Phys. Rep.}, 142 (1986), 103.

\bibitem{Nester_85}
{\sc Lazaridi, A.~N., and Nesterenko, V.~F.}
\newblock Observation of a new type of solitary waves in a one-dimensional
  granular medium.
\newblock {\em J. Appl. Mech. Techn. Phys.}, 26 (1985), 405.

\bibitem{Nester_83}
{\sc Nesterenko, V.~F.}
\newblock Propagation of nonlinear compression pulses in granular media.
\newblock {\em J. Appl. Mech. Techn. Phys.}, 5 (1984), 733.

\bibitem{Nester_94}
{\sc Nesterenko, V.~F.}
\newblock Solitary waves in discrete media with anomalous compressibility and
  similar to "sonic vacuum".
\newblock {\em J. de Physique IV}, 55 (1994), 729.

\bibitem{Nester_02}
{\sc Nesterenko, V.~F.}
\newblock {\em Dynamics of Heterogeneous Materials}.
\newblock Springer-Verlag, New York, 2001.

\bibitem{Nester_95}
{\sc Nesterenko, V.~F., Lazaridi, A.~N., and Sibiryakov, E.~B.}
\newblock The decay of soliton at the contact of two "acoustic vacuums".
\newblock {\em J. Appl. Mech. Techn. Phys.}, 36 (1995), 166.

\bibitem{Vengr}
{\sc Vengrovich, D.~B.}
\newblock private communication.

\bibitem{Vodova}
{\sc Vodova, J.}
\newblock A complete list of conservation laws for non-integrable compacton
  equations of k(n,n) type.
\newblock {\em Nonlinearity}, 26 (2013), 757, (arXiv:1206.440v1 [nlin.SI]).

\end{thebibliography}

\end{document}